\documentclass[pra,showpacs,twocolumn,superscriptaddress,floatfix,aps]{revtex4-2}

\usepackage[dvips]{graphics}
\usepackage{color}
\usepackage{float}
\usepackage{natbib}
\usepackage[dvipsnames]{xcolor}
\setcitestyle{numbers}
\setcitestyle{square}
\usepackage{graphicx}
\usepackage{bm}
\usepackage{amssymb}
\usepackage{amsmath}
\usepackage{mathtools}
\usepackage{dcolumn}  
\usepackage[colorlinks, linkcolor=Blue,citecolor=Blue,urlcolor=Blue]{hyperref}
\usepackage[all]{hypcap} 
\usepackage{physics}
\usepackage{enumerate}
\usepackage{braket}       
\usepackage{overpic}
\usepackage{amsfonts}
\usepackage{dsfont}
\usepackage[mathscr]{eucal}
\usepackage{hyperref}
\usepackage{setspace}

\usepackage{pdfpages} 
\usepackage{pgffor}   

\makeatletter
\AtBeginDocument{\let\LS@rot\@undefined}
\makeatother

\def\supplementfilename{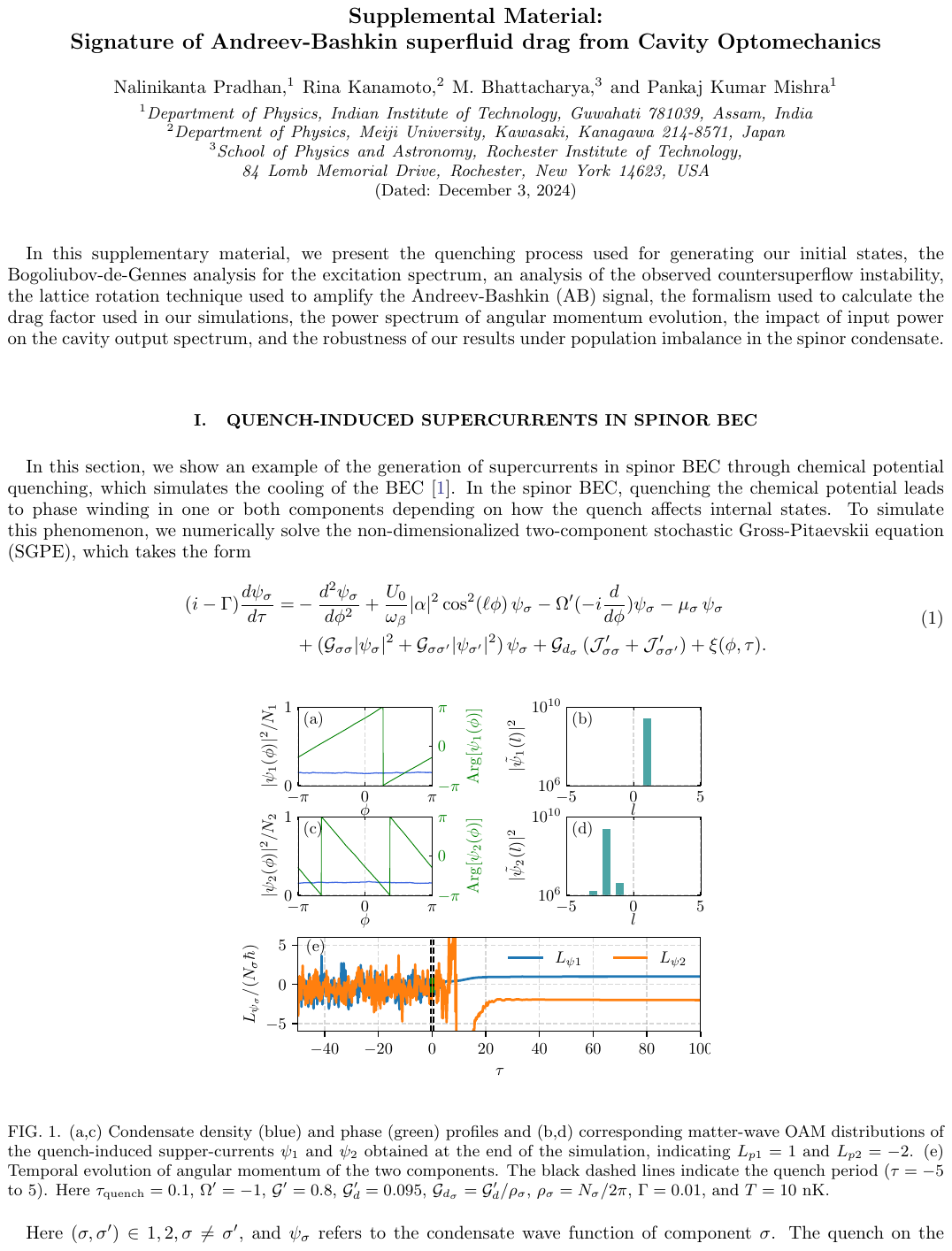}

\pdfximage{\supplementfilename}
\def\numbersupplementpages{\the\pdflastximagepages}

\newif\ifarXiv
\arXivtrue 

\begin{document}
\title{Signature of Andreev-Bashkin superfluid drag from Cavity Optomechanics }
\author{Nalinikanta Pradhan}
\affiliation{Department of Physics, Indian Institute of Technology, Guwahati 781039, Assam, India}

\author{Rina Kanamoto}
\affiliation{Department of Physics, Meiji University, Kawasaki, Kanagawa 214-8571, Japan}

\author{M. Bhattacharya}
\affiliation{School of Physics and Astronomy, Rochester Institute of Technology, 84 Lomb Memorial Drive, Rochester, New York 14623, USA}
\author{Pankaj Kumar Mishra}
\affiliation{Department of Physics, Indian Institute of Technology, Guwahati 781039, Assam, India}

\date{\today}

\begin{abstract} 
The Andreev-Bashkin (AB) effect, corresponding to the dissipationless dragging of one superfluid by another, was predicted almost fifty years ago but has so far eluded experimental detection. In this work, we theoretically introduce an entirely new detection paradigm to this quest, and show that it enables the observation of the hitherto undetected AB effect for realistic parameters. We accomplish this by using the powerful techniques of cavity optomechanics, which were crucial to the observation of gravitational waves, on a spinor ring Bose-Einstein condensate. In contrast to all known AB detection methods, our scheme allows for real-time, \textit{in situ}, minimally destructive and three orders-of-magnitude more sensitive measurement of the AB effect. Our proposal, which considers persistent currents in weakly repulsive atomic condensates, and amplifies the AB signal using a novel dynamic Bragg spectroscopy technique, is supported by numerical simulations of the stochastic Gross-Pitaevski equation, which agree very well with our analytic Bogoliubov-de Gennes calculations. Our work suggests a novel tool for sensitively and nondestructively probing the dynamics of rotationally interacting superfluids using cavities and has fundamental implications for ongoing studies of superfluid hydrodynamics, atomtronics, matter-wave interferometry, and cavity optomechanical sensing.
\end{abstract}

\flushbottom

\maketitle
\textit{Introduction}: The Andreev-Bashkin (AB) effect is a manifestation of macroscopic quantum coherence and describes nondissipative drag exerted by one superfluid on another. Originally considered for the $^{3}$He -$^{4}$He system, where its observation is hindered by limited miscibility   \cite{andreev1975three}, ensuing predictions were made for superconductors \cite{DuanPRLCoulombdrag}, neutron stars \cite{chamel2017superfluidity}, and polaritonic systems \cite{aminov2022superfluid}, but this fundamental effect has eluded observation for almost fifty years. 

With the subsequent availability of highly pure and controllable cold atomic superfluids \cite{PethickBECBook,PoloQST2024}, there has been an intense revival of interest in observing the AB effect \cite{FilPRA2005Two,NespoloNJP2017AB,OtaPRA2020Thermo}, with signatures expected from vortex formation \cite{DahlPRL2008ABVortex,KarlePRA2019BKT}, collective mode frequencies \cite{RomitoPRR2021SpinDipole}, spin wave dynamics \cite{NespoloNJP2017AB,ParisiPRL2018SpinAB, CarliniPRA2021,ContessiPRR2021,SekinoPRR2023SpinC}, transport in optical lattices \cite{LinderPRA2009ABLattice,HoferPRA2012ABLattice,SellinPRB2018ABLattice,SyrwidPRL2021,BlomquistPRL2021Borromean,ContessiPRR2021ABHubbard} and droplets \cite{PylakPRR2022ABDroplet}. Special attention has been attached to geometries that are naturally suited for studies of atomic superflow, i.e. Bose-Einstein condensates (BECs) in ring traps \cite{AmicoRMP2022,ParisiPRL2018SpinAB,ContessiPRR2021, AmicoReview2024}. In these systems, the challenges to be overcome for detection of the AB effect include monitoring of Josephson junctions \cite{FilPRA2005Two}, engineering and detection of soliton dynamics \cite{syrwid2022drag}, operation in the strong coupling regime of atomic interactions \cite{ParisiPRL2018SpinAB}, species selective addressal for rotation and observation, and maximization of the time-of-flight to improve signal contrast \cite{HossainPRA2022Entrainment}. We note that all existing and proposed techniques are fully destructive of the superfluids, as they involve optical absorption imaging.

\begin{figure}[t]
\begin{center}
	\includegraphics[width= 1\linewidth]{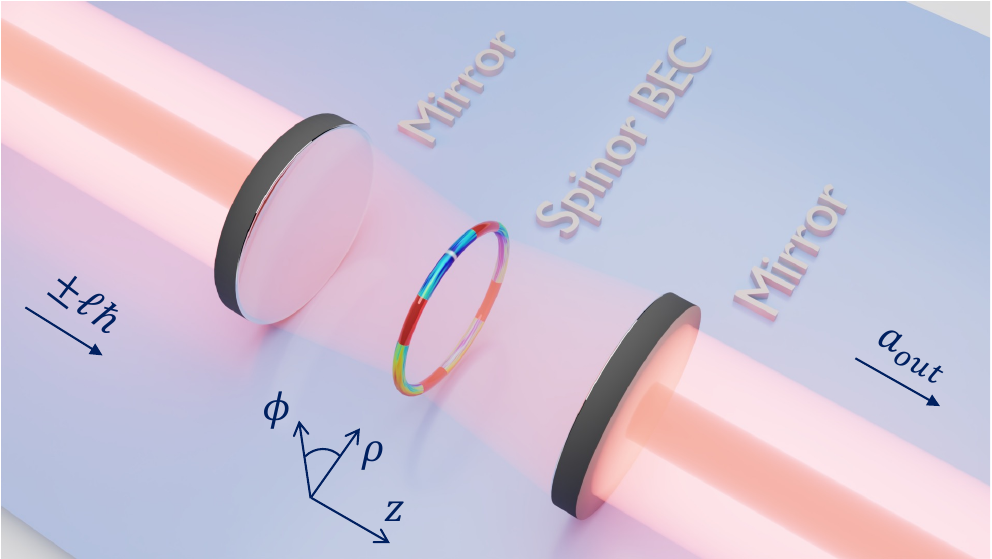} 

\end{center}
\caption{Proposed experimental setup. A spinor BEC is confined in a ring trap inside an optical cavity. The cavity is driven by OAM-carrying optical beams which form an angular lattice on the ring.}
\label{fig:setup}
\end{figure}

In this paper, we theoretically propose a real-time, \textit{in situ}, minimally destructive and resolution-amplified detection of the AB effect, which is three orders of magnitude more sensitive than existing methods and proposals. As shown in Fig.~\ref{fig:setup}, our proposal involves a spinor BEC \cite{CominottiPRL128MassBEC} confined to a ring trap \cite{BeattiePRL2013} inside a cavity driven by beams carrying optical orbital angular momentum (OAM) \cite{KumarPRL2021,pradhan2024ring,pradhan2024cavity}. As demonstrated below, our method does not require Josephson junctions, soliton engineering, strong coupling, species selective addressal, time-of-flight expansion or absorption imaging. Further, it gives direct access, via the optical cavity transmission, to the dynamics of OAM exchange—arising from the AB effect—between the two components of the superfluid BEC. Finally, it uses a novel dynamic Bragg spectroscopy technique to increase the resolution of the AB detection. 

\textit{Theoretical Model}: The central feature of the AB effect in a two-species mixture $[(\sigma,{\sigma'}) \in {1,2}, \sigma\neq {\sigma'}]$ is that the superfluid current $J_\sigma$ of one component depends on the superfluid velocity $v_{\sigma'}$ of the other component \cite{andreev1975three,FilPRA2005Two,NespoloNJP2017AB}. In a two-component BEC, confined in a ring trap, this effect can be modeled by the one-dimensional quantum hydrodynamic Hamiltonian \cite{NespoloNJP2017AB,KumarPRL2021,syrwid2022drag,pradhan2024cavity}
\begin{align}
    \hat{H} &= \sum_\sigma \; \int_{-\pi}^{\pi} \Bigl\{  \Psi_\sigma^{\dagger} (\phi) \left[\frac{\hbar^2}{2m R^2}\left(-i\frac{d}{d\phi}-\frac{\Omega'}{2}\right)^2 \right]\Psi_\sigma(\phi) \nonumber\\
    & + \Psi_\sigma^{\dagger} (\phi)\left[ \hbar U_0 \cos^2 (\ell\phi) a^{\dagger}a\right] \Psi_\sigma(\phi) +  \frac{{g_{\sigma \sigma}}}{2}\; \Psi_\sigma^{\dagger}(\phi)\Psi_\sigma^{\dagger}(\phi)  \nonumber\\
    & \Psi_\sigma(\phi)\Psi_\sigma(\phi) +  \frac{{g_{\sigma {\sigma'}}}}{2}\; \Psi_\sigma^{\dagger}(\phi)\Psi_{\sigma'}^{\dagger}(\phi)\Psi_{\sigma'}(\phi)\Psi_\sigma(\phi)  \nonumber\\
    & + g_{d_{\sigma \sigma}} m \, J_\sigma^2 /2 + \; g_{d_{\sigma {\sigma'}}} m \, J_\sigma J_{\sigma'} \Bigr\} \; d\phi  \nonumber\\ & - \Delta_{0} a^{\dagger}a -i\eta (a-a^{\dagger})\;,\label{Eq:Hamil}
\end{align}
where $\Psi_{\sigma,\sigma'}$ are the bosonic atomic field operators that obey the commutation relation $[\Psi_{\chi} (\phi), \Psi_{\chi} (\phi')] = \delta (\phi - \phi')$, where $\chi \in {\sigma},{\sigma'}$. The first term in Eq.~(\ref{Eq:Hamil}) represents the rotational kinetic energy in the frame co-rotating with the optical lattice, with frequency $\Omega = (\hbar/ 2mR^2)\, \Omega'$, where $m$ is the atomic mass and $R$ is the radius of ring trap (which can be set up, for example, using an optical field resonant with the cavity but far detuned from the atomic transitions). The second term denotes the angular optical lattice potential created on the ring by the superposition of two Laguerre-Gaussian (LG) beams having topological charge $\pm \ell$ respectively \cite{NaidooAPB2012, KumarPRL2021}, orthogonally polarized and detuned from a different atomic transition than the ring trap field. Here $U_0=g_0^{2}/\Delta_a$, where $g_0$ is the single photon-single atom coupling strength and $\Delta_{a}$ is the detuning of the optical lattice from the atomic transition. The photonic creation ($a^{\dagger}$) and annihilation ($a$) operators obey $[a,a^{\dagger} ] = 1$.


The third and fourth terms in  Eq.~(\ref{Eq:Hamil}) represent the intra-species and inter-species density-dependent atomic interactions, with the strengths $g_{{\sigma} {\sigma}}$ and $g_{{\sigma} {\sigma'}}$ respectively. The 1D reduced atomic interaction is $g_{\sigma \chi}= 2\hbar \omega_\rho a_{s_{\sigma \chi}}/R$, where $\omega_\rho$ is the radial harmonic trap frequency and $a_{s_{\sigma \chi}}$ is the $s$-wave atomic scattering length between species $\sigma$ and $\chi$. The fourth-row terms denote the intra-species and inter-species current-dependent atomic interactions with strength $g_{d_{\sigma \chi}}$ \cite{syrwid2022drag}. The probability current density of each component is $J_{\chi} = (\Psi^*_{\chi}\partial_\phi \Psi_{\chi} - \Psi_{\chi}\partial_\Phi \Psi^*_{\chi}) \, \hbar / (2 m R i)$. The first term in the last row represents the cavity field energy in the rotating laser drive frame, with $\Delta_{0}$ being the de-tuning of the driving laser field frequency $\omega_L$ from the cavity resonance frequency $\omega_c$. The last term denotes the coupling of the pump, at the rate $\eta = \sqrt{P_{\mathrm{in}} \gamma_0 / (\hbar \omega_c)} $, where $P_{\mathrm{in}}$ is the input optical power and $\gamma_0$ is the cavity energy decay rate. 

By scaling the energy as $\hbar \omega_{\beta} = \hbar^2 / 2 m R ^2$ and time as $\tau = \omega_\beta t\;$, from the Hamiltonian [Eq. (\ref{Eq:Hamil})], and using quantum noise theory \cite{KumarPRL2021}, the coupled dimensionless stochastic Gross-Pitaevskii equations (sGPE) for the condensate order parameters $\psi_{\sigma}$ (classical wave functions equivalent to the atomic operators $\Psi_\sigma$ in the Gross-Pitaevski approximation \cite{PethickBECBook}) and the light field amplitude $\alpha$ can be written as \cite{syrwid2022drag,pradhan2024cavity,pradhan2024ring}
\begin{equation}
\begin{aligned}
    (i-\Gamma)\frac{d \psi_{\sigma}}{d\tau} = & -\frac{d^2 \psi_{\sigma}}{d\phi ^2} + \frac{U_0}{\omega_\beta} |\alpha|^2 \cos^2(\ell \phi) \, \psi_\sigma - \mu_\sigma \, \psi_\sigma  \\ &-\Omega' (-i \frac{d}{d \phi}) \psi_\sigma + (\mathcal{G}_{\sigma \sigma}  |\psi_{\sigma}|^2 + \mathcal{G}_{\sigma {\sigma'}}  |\psi_{\sigma'}|^2 ) \, \psi_{\sigma}   \\ &+ \mathcal{G}_d \, (\mathcal{J}'_{\sigma \sigma} +  \mathcal{J}'_{\sigma {\sigma'}}) + \xi(\phi,\tau),
    \label{Eq:bec}    
\end{aligned}
\end{equation}
and
\begin{equation}
\begin{aligned}
    i\frac{d\alpha}{d\tau} = \biggl\{ - \biggl[\Delta_{0} &- U_0 \langle {\cos^2\left({\ell\phi}\right)}\rangle_{\tau} + i \frac{\gamma_{0}}{2}\biggr] \alpha  + i\eta \biggr\} \omega_\beta^{-1}\\ & + i \sqrt{\gamma_0} \omega_\beta^{-1} \alpha_{in}(\tau).
    \label{Eq:cavity}
\end{aligned}
\end{equation}
Here each condensate wave function $\psi_\sigma$ is normalized to the number of atoms ($N$) as $\int_{-\pi}^{\pi} |\psi_\sigma(\phi,\tau)|^2 \, d\phi = N$. 

The first term on the right-hand side of the first row of Eq.~(\ref{Eq:bec}) represents the scaled kinetic energy, the second term is the light-matter interaction, and the last term denotes the scaled chemical potential, which serves to conserve the norm by being corrected at each time step as~\cite{mithun2018signatures} $\Delta\mu_\sigma = (\Delta \tau)^{-1} \ln{\left[\int \lvert \psi_\sigma(\phi, \tau)\rvert^2 d\phi / \int \vert\psi_\sigma(\phi, \tau + \Delta \tau)\rvert^2 d\phi\right]}$. The second term of the second row denotes the scaled intra-species and inter-species density-dependent atomic interactions with strengths $\mathcal{G}_{\sigma \sigma}=g_{\sigma \sigma} / (\hbar \omega_\beta)$ and $\mathcal{G}_{\sigma {\sigma'}}=g_{\sigma {\sigma'}} / (\hbar \omega_\beta)$ respectively and we use $ \mathcal{G}' = \mathcal{G}_{\sigma {\sigma'}} / \mathcal{G}_{\sigma \sigma} $. In the third row, $\mathcal{J}'_{\sigma \sigma}$ and $\mathcal{J}'_{{\sigma} {\sigma'}}$ represent the scaled intra-species and inter-species current-current interaction terms. Here, $\mathcal{J}'_{\sigma \chi}= [2 (\partial_\phi \psi_{\sigma}) J'_{\chi} + \psi_{\sigma}(\partial_\phi J'_{\chi})]\, /2i$ and $J'_{\chi}$ is the scaled probability current density, expressed as $J'_{\chi} = (\psi^*_{\chi}\partial_\phi \psi_{\chi} - \psi_{\chi}\partial_\Phi \psi^*_{\chi})/(2 i)$.  The strength of these interactions is given by $\mathcal{G}_d = \mathcal{G}_d' / \rho_\sigma$, where $\rho_\sigma = N/ 2 \pi$ is the condensate density. By comparing the drag energy density in Eq.(~\ref{Eq:Hamil}) with the free energy density of a mixture of two superfluids, we obtain the drag parameter $\mathcal{G}_d'  = 2 g_{d_{\sigma \chi}} \rho_1 =  2 \rho_{d} / \rho_1 $, where $\rho_1 = \rho_2$ is the density of each component and $\rho_d$ is the drag density arising due to the current coupling between two components~\cite{ParisiPRL2018SpinAB}. The parameter $\mathcal{G}_d'$ depends on the inter-species and intra-species density-density interactions \cite{ParisiPRL2018SpinAB}, with more details on the appropriate range of values of $\mathcal{G}_d'$ given in the supplementary material~\cite{supplementary}.

The classical mean-field dynamics of the intracavity light field amplitude $\alpha$ is governed by Eq.~(\ref{Eq:cavity}). The second term within the square bracket of the right-hand side of Eq.~(\ref{Eq:cavity}) represents the coupling between the light mode and condensates, where the expectation value of the optical lattice potential is given by $ \langle {\cos^2\left({\ell\phi}\right)}\rangle_{\tau} =  \sum_{\sigma}\int_{-\pi}^{\pi} |\psi_\sigma(\phi,\tau)|^2 \cos^2\left({\ell\phi}\right) d\phi$ \cite{pradhan2024ring,pradhan2024cavity}. The last terms of Eqs.~(\ref{Eq:bec}) and (\ref{Eq:cavity}) represent the thermal fluctuations $\xi(\phi,\tau)$ present in the condensate and the shot noise $\alpha_{in}(\tau)$ carried by the laser light, respectively. Both are delta-correlated white noises having zero mean and unit variance with the correlations \cite{das2012winding, pradhan2024cavity}
\begin{align} 
\langle\xi(\phi,\tau) \, \xi ^ *(\phi',\tau')\rangle  &= \frac{2\,\Gamma \,k_B \,T}{\hbar \omega_\beta} \, \delta(\phi - \phi') \, \delta (\tau - \tau'), \\ 
\langle \alpha_{in}(\tau) \, \alpha_{in}^ *(\tau') \rangle  &= \omega_\beta \, \delta (\tau - \tau'),
\end{align}
where $\Gamma$ and $T$ denote the dissipation and temperature of the condensate, respectively, and $k_B$ is the Boltzmann constant.

We solve Eqs.~(\ref{Eq:bec}) and (\ref{Eq:cavity}) in real-time using a fourth-order Runge-Kutta scheme \cite{tan2012general}. In our simulations, we consider $^{23}$Na atoms in the two Zeeman states $|F,M_{F}\rangle=|1,\pm 1\rangle$, where $F(M_{F})$ is the quantum number of atomic hyperfine (magnetic) angular momentum quantum number; the two states are rendered non-degenerate by a small uniform magnetic field \cite{CominottiPRL128MassBEC}. The OAM-carrying beams are detuned to the $^{23}$Na D$_{2}$ transition at $589$ nm. We rotate the optical lattice at frequency $\Omega$ (this may be accomplished experimentally using spatial light modulators) such that $\cos^{2}(l\phi) \rightarrow \cos^{2}(l\phi+\Omega't)$ in Eqs.~(\ref{Eq:bec}) and (\ref{Eq:cavity}), where $\Omega'=\Omega/\omega_{\beta}$ \cite{pradhan2024ring}. The parameter values used are $m = 23$ amu, $R = 12$ $\mu$m, $\Gamma = 0.0001$, $T = 10$ nK, $ U_0 = 2\pi \times 212$ Hz, $\Tilde{\Delta} = \Delta_0 - U_0 N = -2 \pi \times 173$ Hz, $\Delta_a = 2\pi \times 4.7$ GHz, $\omega_c = 2\pi \times10^{15} $ Hz, $\omega_\rho = \omega_z = 2\pi \times 42$ Hz, $\gamma_{0} = 2\pi\times 2$ MHz and $\Omega' = -1$ \cite{WrightPRL2013, EckelNature2014}. For the initial calculations, we have used the intraspecies scattering length $a_{s_{\sigma\sigma}} = 2.5$ nm and the inter-species scattering length $a_{s_{\sigma \sigma'}} =  \mathcal{G}' a_{s_{\sigma\sigma}}$. Later in the paper, we address variations in the interaction strengths, considering that the various relevant $^{23}$Na ground state Zeeman states display nearly the same scattering lengths \cite{KnoopPRA2011}.   


\begin{figure}[b]
\begin{center}
	\includegraphics[width= 1\linewidth]{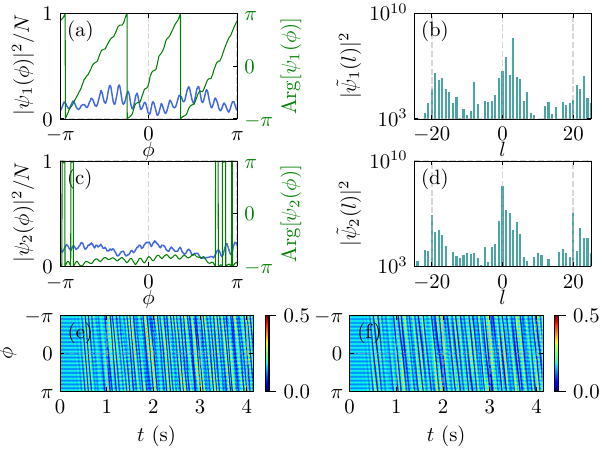} 

\end{center}
\caption{(a,c) Condensate density (blue), phase (green) profiles, and (b,d) matter-wave OAM distributions of the two components $\psi_1$ and $\psi_2$ for $L_{p_1} = 3, L_{p_2} = 0$ after $4$ seconds of time evolution. The other parameters are $\ell = 10$, $N =  1500$, $\Omega' = -1$, $\mathcal{G}' = 0.8$,  $\mathcal{G}_d' = 0.095$, $P_{\mathrm{in}} = 0.4 \, \mathrm{pW}$. (e,f) Pseudo-color plots showing the temporal evolution of the condensate density profile of $\psi_1$ and $\psi_2$, scaled by $N$.}
\label{fig:fig2}
\end{figure}


\textit{Results}: The initial wave functions, assumed by us are
\begin{equation}
    \psi_\sigma(\phi, 0) =  \sqrt{\frac{N}{2 \pi}} e^{i L_{p_\sigma} \phi},
\end{equation}
where $\sigma \in {1,2}, $ and $L_{p_1}$ and $L_{p_2}$ are the winding numbers of the two components, respectively. We consider one component initially rotating ($L_{p_1} \neq 0$) in the presence of a second, initially non-rotating $L_{p_2} = 0$ condensate. We have numerically verified that such supercurrent distributions are experimentally feasible via a rapid temperature quench \cite{weiler2008spontaneous, das2012winding} or by merging independent condensates in a ring trap \cite{aidelsburger2017relaxation}; an example is provided in the supplementary material \cite{supplementary}. The AB drag effect facilitates the transfer of angular momentum from the initially rotating to the non-rotating component \cite{andreev1975three,syrwid2022drag}.

\begin{figure}[b]
\includegraphics[width=1\linewidth]{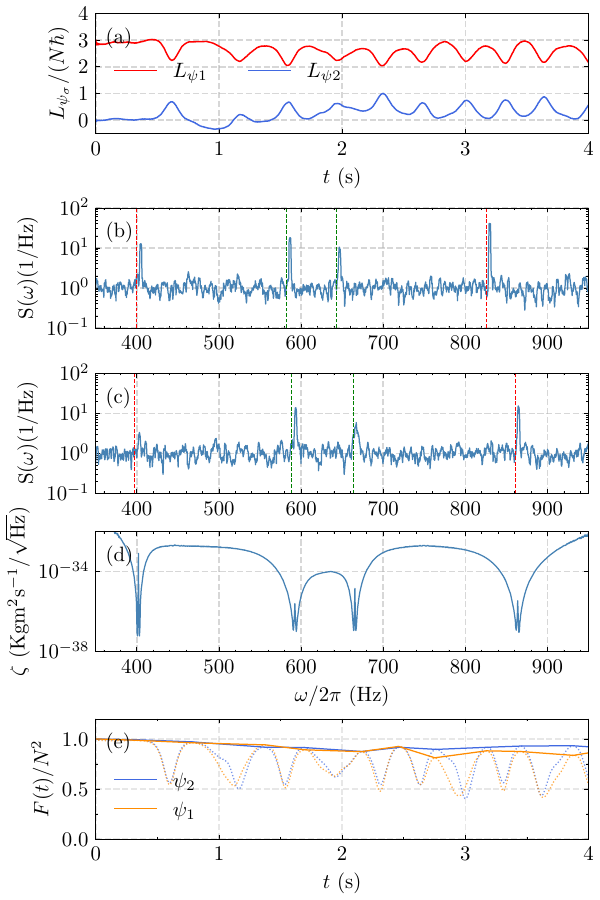} \\
\caption{(a) temporal evolution of the angular momenta per atom of the two components. (b) Power spectra of the output phase quadrature of the cavity field in the absence of current-current coupling ($\mathcal{G}_d' = 0$, $\mathcal{G}' = 0.8$) and (c) in the presence of current-current coupling $(\mathcal{G}_d' = 0.095$, $\mathcal{G}' = 0.8)$. The red (green) vertical dashed lines in (a) and (b) denote the analytical predictions for the side mode frequencies corresponding to  $L_{p_1} \pm 2\ell- \Omega'/2$ ($L_{p_2} \pm 2\ell- \Omega'/2$) obtained from the BdG analysis. (d) Rotation measurement sensitivity versus system response frequency. Here $\Omega' = -1$, $G = 2 \pi \times 2.90$ kHz, $|\alpha_s|^2 = 0.19$, and $t_{meas} = 24.6$ ms.  (e) Variation of fidelity with time. The other parameters used are the same as in Fig.~\ref{fig:fig2}. }
  \label{fig:fig3}
\end{figure}

\begin{figure}[b]
\includegraphics[width=1\linewidth]{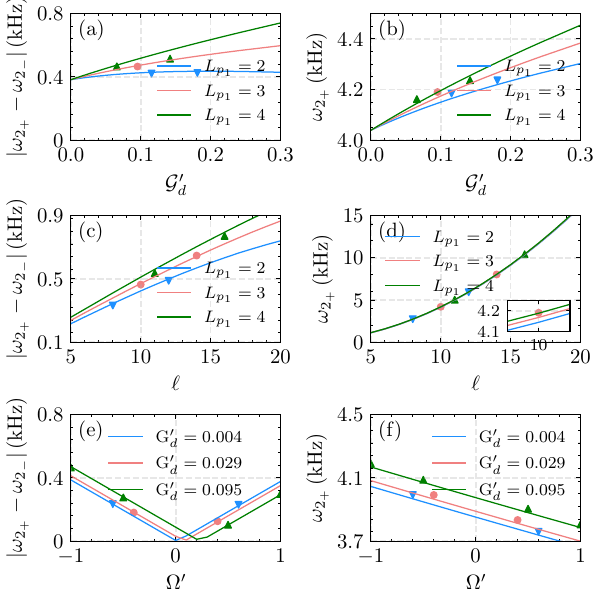} \\
\caption{Variation of the peak splitting $|\omega_{2_+} - \omega_{2_-}|$ and the peak location of $\omega_{2+}$ (a,b) with $\mathcal{G}_d'$ for $\ell = 10$, $\Omega' = -1$, (c,d) with $\ell$ for $\mathcal{G}_d' = 0 .095 \, ( \mathcal{G}' = 0.8)$, $\Omega' = -1$, and (e,f) with $\Omega'$ for $L_{p_1} = 3$ and $\ell = 10$. The other parameters used are the same as in Fig.~\ref{fig:fig2}. The data points denoted by markers are obtained from the numerical simulations, whereas the lines are obtained from the Bogoliubov analysis.}
  \label{fig:fig4}
\end{figure}

With these initial conditions, we simulate the system up to $t = 4$ seconds, well within the demonstrated lifetime of supercurrents, which is of the order of minutes \cite{beattie2013persistent,GuoPRL2020}. The results of these simulations are shown in Figs.~\ref{fig:fig2},
~\ref{fig:fig3} and ~\ref{fig:fig4}.  The condensate density (blue) and phase (green) profiles of $\psi_1$ and $\psi_2$ are shown in Figs.~\ref{fig:fig2} (a) and (c). The phase profile of $\psi_{1}$ validates the winding number $L_{p_1} = 3$, while the phase profile of $\psi_{2}$, which was initially non-rotating, shows a blunted jump, indicating the presence of induced rotation in the second component. The high-frequency modulations in the condensate densities and phases are due to the weak optical lattice on the ring. 

To assess the rotational states of the two condensates, we show the matter wave OAM distributions ($\lvert\tilde{\psi_\sigma}(l)\rvert^2$) in Figs.~\ref{fig:fig2} (b) and (d). Here $\tilde{\psi}(l)$ is the Fourier transformation, taken in the frame of the rotating lattice for convenience, of the condensate wave function $\psi(\phi)$ into the matter wave OAM basis. Due to Bragg scattering of the atoms from the optical lattice, a fraction of atoms undergo a change in their winding number from $L_{p_\sigma}$ to $L_{p_\sigma} \pm 2\ell$ (since the photon number inside the cavity is less than one on average, the lattice is weak and higher-order diffraction is negligible \cite{BrenneckeScience2008,KumarPRL2021}). The significant occupancy of the modes adjacent to $L_{p_\sigma}$ and $ L_{p_\sigma} \pm 2\ell$ in Figs.~\ref{fig:fig2} (b) and (d) indicates that the plane wave states are modulated to a solitonic state due to the transfer of angular momentum from rotating to non-rotating atoms \cite{syrwid2022drag}. (Few-mode plane wave versus multi-mode solitonic matter wave OAM distributions can be seen in our earlier work \cite{pradhan2024cavity}). The real-time evolution of the density profiles is shown in Figs.~\ref{fig:fig2} (e) and (f), where the visible stripes reveal the dynamical formation of matter-wave solitons due to the superfluid drag effect and their subsequent motion in the ring trap \cite{syrwid2022drag}.


We compute the angular momentum per atom of both superfluid components to demonstrate the drag. Fig.~\ref{fig:fig3} (a) shows a repeated exchange of angular momentum as time evolves \cite{syrwid2022drag}. 
In our proposal, this exchange can be optically detected via homodyne measurement of cavity transmission \cite{KumarPRL2021,pradhan2024cavity}. The relevant theoretical quantity is the power spectrum of the phase quadrature of the cavity output field, defined as 
\begin{equation}
\label{eq:Spect}
S(\omega)=\left|\mathrm{Im}\left[\alpha_{out}(\omega)\right]\right|^{2},
\end{equation}
where the cavity output field $\alpha_{out} = -\alpha_{in} + \sqrt{\gamma_0}\alpha$ is related to the input $(\alpha_{in})$ and intracavity $(\alpha)$ fields  \cite{AspelmeyerRMP2014}. Spectra, shown in Figs.~\ref{fig:fig3}(b) and (c) have well-resolved peaks at $\omega_{1\pm}$ (outer peaks) and $\omega_{2\pm}$ (inner peaks), corresponding to the matter wave sidemodes with winding numbers $L_{p_1}  \pm 2\ell- \Omega' / 2$ and $L_{p_2}  \pm 2\ell- \Omega' / 2$, respectively, where $I=mR^{2}$ is the moment of inertia of each atom about the cavity axis. Identification of individual peaks is made possible by performing a Bogoliubov analysis (see details in supplementary material~\cite{supplementary} ), and the frequencies obtained are indicated by vertical dashed lines in all cavity output spectra in Fig.~\ref{fig:fig3}. 

We now examine how the cavity output spectra behave as the interactions in the system vary, an analysis crucial for identifying the AB effect. In the absence of \textit{all} inter-species interactions (density-density as well as current-current) and no lattice rotation $(\Omega'=0)$, the initially nonrotating second component remains nonrotating ($L_{p_2} = 0$), and the $\omega_{2\pm}$ Bragg peaks are degenerate. Introducing \textit{only} the density-density inter-component interactions facilitates angular momentum transfer via counter superflow instability, as explained in \cite{TakeuchiPRL2010,law2001critical}, and the supplementary material \cite{supplementary}. It leads to splitting of the $\omega_{2\pm}$ peaks, which, however, can only be resolved by rotating the lattice $(\Omega'\neq 0)$ as shown in Fig.~\ref{fig:fig3}(b). Finally, by including the current-current interactions, which introduce the AB effect \cite{NespoloNJP2017AB,syrwid2022drag}, further splitting of the $\omega_{2\pm}$ peaks is observed as can be seen in Fig.~\ref{fig:fig3} (c). This additional splitting is a clear signature of the AB effect and is the central result of this paper. 


The effectiveness of our proposed method can be estimated by calculating its sensitivity to rotation measurements, which is defined as \cite{KumarPRL2021}
\begin{equation}
\label{eq:Sensitivity}
    \zeta = \frac{S(\omega)}{\partial S(\omega)/\partial\Lambda} \times \sqrt{t_{meas}},
\end{equation}
where $\Lambda = \hbar L_{p_2}$ is the angular momentum of the initially non-rotating condensate. Here $t_{meas}^{-1} = 8 (\alpha_s G)^2 / \gamma_0$ is the optomechanical measurement rate in the bad cavity limit, $\alpha_s$ is the steady state of the cavity field, and $G = U_0 \sqrt{N} / 2 \sqrt{2}$ \cite{AspelmeyerRMP2014,KumarPRL2021}. The sensitivity of rotation measurement on the initially non-rotating component is shown as a function of the system response frequency in Fig.~\ref{fig:fig3} (d). The best sensitivities of $(\sim 10^{-37}$kg m$^{2}$s$^{-1}$/$\sqrt{\mathrm{Hz}})$, obtained near the side-mode frequencies of the condensate, are about three orders of magnitude better than existing methods and schemes, which are all based on optical absorption imaging \cite{KumarNJP2016, KumarPRL2021}. We note that for our system, $t_{meas} (\sim 24.6 \,\mathrm{ms}) \ll \Gamma^{-1}(\sim 60\,\mathrm{s})$, which means the proposed measurement can be carried out practically in real-time before the BEC rotation is damped significantly. 

To characterize the extent to which the superfluid components are disturbed by the rotation measurement, we have also calculated the fidelity $F(t) = \int_{-\pi}^{\pi} \left[\psi_\sigma^{*}(\phi,t)\psi_\sigma(\phi,0)\right]^2 d\phi$ of the wave functions over time, as shown in Fig.~\ref{fig:fig3} (e). As can be seen, the fidelity stays around $0.9$ over $4$ seconds. From detailed analyses, we have found that the decrease in fidelity is more due to the transition from the uniform density condensate to the solitonic state (as part of the OAM transfer mechanism) and less due to the rotation measurement (i.e., interaction with the light field). This confirms the minimally destructive nature of our measurement technique.

We now analyze the variation of the peak locations and separations on relevant experimental parameters, as shown in Fig.~\ref{fig:fig4}. This allows us to account for experimental parameter variations and to optimize configurations where possible. In the figure, the solid lines are from the Bogoliubov analysis, and the markers denote numerical simulation data. As can be seen in Fig.~\ref{fig:fig4} (a) the peak splitting $(|\omega_{2+} - \omega_{2-}|)$ increases with current-current interaction between species $\mathcal{G}_d'$, which can be modified by the choice of different initial Zeeman states via the scattering length $a_{s_{\sigma \sigma'}}$ \cite{KnoopPRA2011}. The variation in the peak splitting of the peak on the topological charge $\ell$ of the LG beam is shown in Fig.~\ref{fig:fig4}(c). The location of the $\omega_{2+}$ peak is shown in Fig.~\ref{fig:fig4} (b) and (d) as a function of $\mathcal{G}_d'$ and $\ell$, respectively. The variation of the peak splitting and $\omega_{2+}$ peak location with the lattice rotation ($\Omega'$) is shown in Fig.~\ref{fig:fig4} (e) and (f). In all cases, the simulations and the Bogoliubov analysis agree well. These results provide a template for mapping experimental data and determining the AB drag factor of the two interacting superfluids. Before concluding, we would like to emphasize that while all figures in this paper relate to the physics of the system, only the cavity transmission spectra (and ensuing sensitivity, splittings, etc.) can be measured with minimal destruction. The condensate density and phase profile measurements, for example, would result in the complete destruction of the condensates and have only been provided to clarify the physics.

\textit{Conclusion}: We have theoretically introduced an entirely new detection paradigm for observing the heretofore undetected Andreev-Bashkin (AB) effect, which predicts dissipationless drag to be exerted by one superfluid upon another. In contrast to all other AB detection methods in existence, our proposal allows for real-time, \textit{in situ}, minimally destructive, resolution-amplified and three orders-of-magnitude more sensitive measurement of the AB effect using experimentally available spinor BECs in ring traps and standard optical cavities and beams. Our theoretical analysis is based on numerical simulations of the stochastic Gross-Pitaevski equation as well as analytic Bogoliubov theory, which show good agreement with each other. In our proposal, we have presented condensate density and phase profiles, matter-wave orbital angular momentum spectra, cavity output spectra, and measurement sensitivities and fidelities. Our proposal brings for the first time the powerful techniques of cavity optomechanics, which enabled the detection of gravitational waves, to the study of rotationally interacting superfluids.  We expect our work to have fundamental implications for ongoing studies of superfluid hydrodynamics, atomtronics, matter-wave interferometry, cosmological simulations, and cavity optomechanical sensing.

M.B. thanks the Air Force Office of Scientific Research (FA9550-23-1-0259) for support. R.K. acknowledges support from JSPS KAKENHI Grant No. JP21K03421. We also gratefully acknowledge our supercomputing facility Param-Ishan (IITG), where all the simulation runs were performed.

\bibliography{citation.bib} 

\ifarXiv


\section*{Supplementary material}

# Supplemental Material:
# Signature of Andreev-Bashkin superfluid drag from Cavity Optomechanics

Nalinikanta Pradhan,[1] Rina Kanamoto,[2] M. Bhattacharya,[3] and Pankaj Kumar Mishra[1]

[1]*Department of Physics, Indian Institute of Technology, Guwahati 781039, Assam, India*
[2]*Department of Physics, Meiji University, Kawasaki, Kanagawa 214-8571, Japan*
[3]*School of Physics and Astronomy, Rochester Institute of Technology,
84 Lomb Memorial Drive, Rochester, New York 14623, USA*
(Dated: December 3, 2024)

In this supplementary material, we present the quenching process used for generating our initial states, the Bogoliubov-de-Gennes analysis for the excitation spectrum, an analysis of the observed countersuperflow instability, the lattice rotation technique used to amplify the Andreev-Bashkin (AB) signal, the formalism used to calculate the drag factor used in our simulations, the power spectrum of angular momentum evolution, the impact of input power on the cavity output spectrum, and the robustness of our results under population imbalance in the spinor condensate.

## I. QUENCH-INDUCED SUPERCURRENTS IN SPINOR BEC

In this section, we show an example of the generation of supercurrents in spinor BEC through chemical potential quenching, which simulates the cooling of the BEC [1]. In the spinor BEC, quenching the chemical potential leads to phase winding in one or both components depending on how the quench affects internal states. To simulate this phenomenon, we numerically solve the non-dimensionalized two-component stochastic Gross-Pitaevskii equation (SGPE), which takes the form

$$(i-\Gamma)\frac{d\psi_\sigma}{d\tau} = -\frac{d^2\psi_\sigma}{d\phi^2} + \frac{U_0}{\omega_\beta}|\alpha|^2\cos^2(\ell\phi)\,\psi_\sigma - \Omega'(-i\frac{d}{d\phi})\psi_\sigma - \mu_\sigma\,\psi_\sigma \\ + (\mathcal{G}_{\sigma\sigma}|\psi_\sigma|^2 + \mathcal{G}_{\sigma\sigma'}|\psi_{\sigma'}|^2)\,\psi_\sigma + \mathcal{G}_{d_\sigma}\,(\mathcal{J}'_{\sigma\sigma} + \mathcal{J}'_{\sigma\sigma'}) + \xi(\phi,\tau). \quad (1)$$

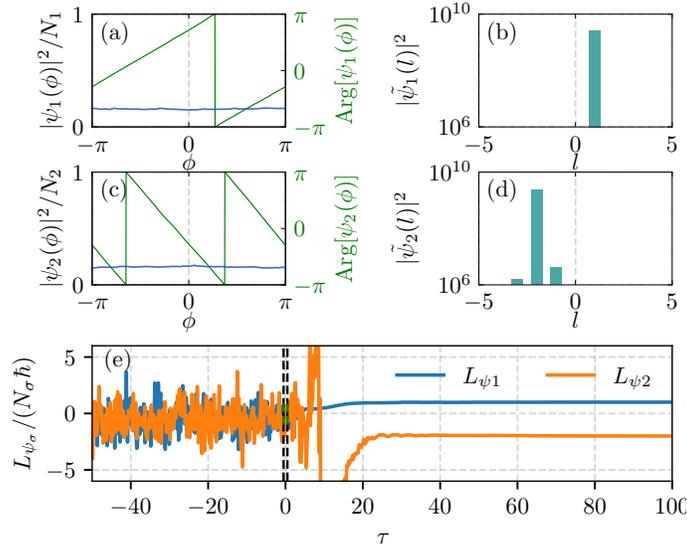

FIG. 1. (a,c) Condensate density (blue) and phase (green) profiles and (b,d) corresponding matter-wave OAM distributions of the quench-induced supper-currents $\psi_1$ and $\psi_2$ obtained at the end of the simulation, indicating $L_{p1}=1$ and $L_{p2}=-2$. (e) Temporal evolution of angular momentum of the two components. The black dashed lines indicate the quench period ($\tau=-5$ to 5). Here $\tau_{\text{quench}}=0.1$, $\Omega'=-1$, $\mathcal{G}'=0.8$, $\mathcal{G}'_d=0.095$, $\mathcal{G}_{d_\sigma}=\mathcal{G}'_d/\rho_\sigma$, $\rho_\sigma=N_\sigma/2\pi$, $\Gamma=0.01$, and $T=10$ nK.

Here $(\sigma,\sigma')\in 1,2, \sigma\neq\sigma'$, and $\psi_\sigma$ refers to the condensate wave function of component $\sigma$. The quench on the

chemical potential follows the following protocol given by [1]

$$\mu(\tau) = \begin{cases} -\mu_0 & \text{for } \tau < -\tau_{\text{quench}}, \\ \mu_0 \frac{\tau}{\tau_{quench}} & \text{for } \tau \in [-\tau_{\text{quench}}, \tau_{\text{quench}}], \\ \mu_0 & \text{for } \tau > \tau_{\text{quench}}. \end{cases} \quad (2)$$

Here $\tau_{\text{quench}} = 5$ and $\mu_0 = 50$. This protocol varies the chemical potential in the range $-\mu_0$ to $\mu_0$ linearly in the time interval $-\tau_{\text{quench}} \leq \tau \leq \tau_{\text{quench}}$.

The condensate density profile, phase profile, and OAM distributions for the two superflow cases are shown in Fig. 1. In panels (a) and (c), the density profiles are scaled by the norm of each component. From the phase profiles and the OAM distribution, it is clear that the $\psi_1$ state has obtained the winding number $L_{p_1} = 1$, whereas for the other component, the winding number $L_{p_2} = -2$. The quench dynamics of the angular momentum of both components throughout $-50 < \tau < 150$ are shown in Fig. 1 (e). This allows the system to stabilize thermally and one of the components to acquire a non-zero winding number. Here it should be noted that, due to the stochastic nature of the system, the winding number of the induced superflows is different for different realizations. Thus we have similarly realized $L_{p1} = 3, L_{p2} = 0$ as used in the main manuscript.

## II. BOGOLIUBOV-DE GENNES ANALYSIS

In this section, we present the Bogoliubov-de-Gennes (BdG) analysis for our model. The extended Gross-Pitaevskii equation, including the current-dependent atomic interactions, takes the form [2–4]

$$i\hbar \frac{d\psi_\sigma}{dt} = -\frac{\hbar^2}{2mR^2}\left[\frac{d^2}{d\phi^2} + \Omega'\left(-i\frac{d}{d\phi}\right)\right]\psi_\sigma + \hbar U_0 |\alpha|^2 \cos^2(\ell\phi)\psi_\sigma \\ + (g_{\sigma\sigma}|\psi_\sigma|^2 + g_{\sigma\sigma'}|\psi_{\sigma'}|^2)\psi_\sigma + g_{d_{\sigma\sigma}} m \mathcal{J}_{\sigma\sigma} + g_{d_{\sigma\sigma'}} m \mathcal{J}_{\sigma\sigma'}, \quad (3)$$

where $\sigma, \sigma' \in (1, 2)$, $\sigma' \neq \sigma$. In this equation, along with the rotational kinetic energy, potential energy due to the optical lattice, and the conventional density-dependent atomic interactions, we have the last two terms on the right-hand side that represent the current-dependent atomic interactions, having the form

$$\mathcal{J}_{\sigma\chi} = \left[2(\partial_\phi \psi_\sigma)(\psi_\chi^* \partial_\phi \psi_\chi - \psi_\chi \partial_\phi \psi_\chi^*) + \psi_\sigma(\psi_\chi^* \partial_\phi^2 \psi_\chi - \psi_\chi \partial_\phi^2 \psi_\chi^*)\right](\hbar/2mRi)^2 \quad (4)$$

where $\chi \in (\sigma, \sigma')$. To obtain the excitation spectrum, we perturb the spatially uniform state, replacing it with $\psi_\sigma = (\Phi_\sigma + \delta\Phi_\sigma)e^{-i\mu_\sigma t/\hbar}$. We obtain the linearized equations for the small perturbation $\delta\Phi_\sigma$ and using

$$\Phi_\sigma = \sqrt{\frac{N}{2\pi}} e^{iL_{p\sigma}\phi} \quad (5)$$

and

$$\delta\Phi_\sigma = e^{iL_{p\sigma}\phi}\left[\frac{u_\sigma}{\sqrt{2\pi}} e^{i(\pm 2\ell\phi - \omega t)} + \frac{v_\sigma^*}{\sqrt{2\pi}} e^{-i(\pm 2\ell\phi - \omega^* t)}\right], \quad (6)$$

we get

$$\hbar\omega \begin{bmatrix} u_1 & v_1 & u_2 & v_2 \end{bmatrix}^T = M \begin{bmatrix} u_1 & v_1 & u_2 & v_2 \end{bmatrix}^T = (M_0 + M_{int} + g_d' M_{g_d}) \begin{bmatrix} u_1 & v_1 & u_2 & v_2 \end{bmatrix}^T. \quad (7)$$

The matrix $M_0$, which includes the contributions from the rotational kinetic energy and the chemical potential is expressed as

$$M_0 = \begin{bmatrix} \hbar\omega_1 - \mu_1 & 0 & 0 & 0 \\ 0 & -\hbar\omega_1 + \mu_1 & 0 & 0 \\ 0 & 0 & \hbar\omega_2 - \mu_2 & 0 \\ 0 & 0 & 0 & -\hbar\omega_2 + \mu_2 \end{bmatrix}, \quad (8)$$

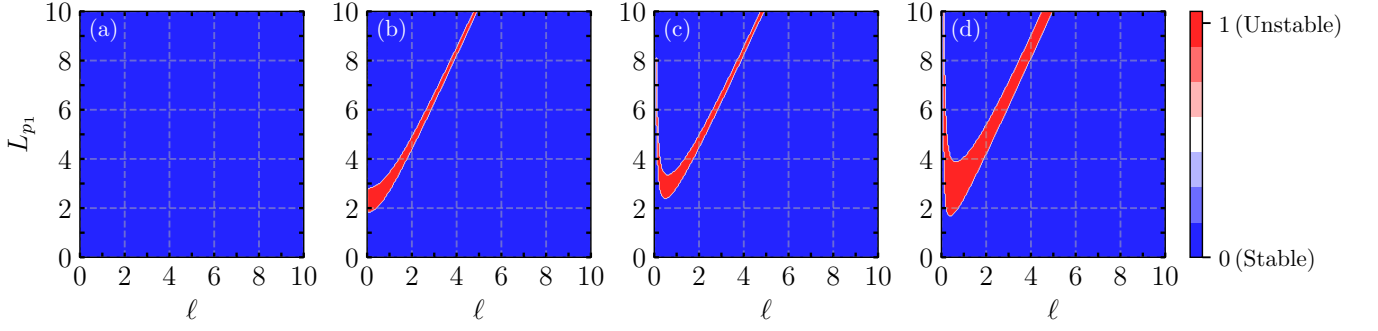

FIG. 2. Stability phase diagram for (a) $\mathcal{G}'_d = 0$ ($\mathcal{G}' = 0$), $\Omega' = 0$, (b) $\mathcal{G}'_d = 0.018$ ($\mathcal{G}' = 0.4$), $\Omega' = 0$, (c) $\mathcal{G}'_d = 0.018$ ($\mathcal{G}' = 0.4$), $\Omega' = -1$, and (d) $\mathcal{G}'_d = 0.095$ ($\mathcal{G}' = 0.8$), $\Omega' = -1$. In all the plots, $L_{p_2} = 0$ and $N = 1500$. The blue (red) region indicates the dynamically stable (unstable) region.

where $\omega_\beta = \hbar/(2mR^2)$, $\omega_{1\pm} = \hbar(L_{p_1} \pm 2\ell - \Omega'/2)^2/2I$, $\omega_{2\pm} = \hbar(L_{p_2} \pm 2\ell - \Omega'/2)^2/2I$, and

$$\mu_\sigma = \hbar[\omega_\beta(L_{p_\sigma}^2 - \Omega' L_{p_\sigma}) + 2N(\tilde{g}_{\sigma\sigma} + \tilde{g}_{\sigma\sigma'})] + \sum_{\chi=\sigma,\sigma'} g_{d_{\sigma\chi}} m\,(\hbar/2mRi)\,(\hbar/2mRi)\,[2(iL_{p_\sigma})(2iL_{p_\chi}(N/2\pi))]. \quad (9)$$

The matrix $M_{int}$ in Eq. (7) represents the density-dependent atomic interaction energy and it is expressed as

$$M_{int} = 2N\hbar \begin{bmatrix} 2\tilde{g}_{11} + \tilde{g}_{12} & \tilde{g}_{11} & \tilde{g}_{12} & \tilde{g}_{12} \\ -\tilde{g}_{11} & -(2\tilde{g}_{11} + \tilde{g}_{12}) & -\tilde{g}_{12} & -\tilde{g}_{12} \\ \tilde{g}_{21} & \tilde{g}_{21} & 2\tilde{g}_{21} + \tilde{g}_{21} & \tilde{g}_{21} \\ -\tilde{g}_{21} & -\tilde{g}_{21} & -\tilde{g}_{21} & -(2\tilde{g}_{21} + \tilde{g}_{21}) \end{bmatrix}, \quad (10)$$

where $\tilde{g}_{\sigma\chi} = g_{\sigma\chi}/(4\pi\hbar)$. The last term in Eq. (7) represents the current-current coupling and it is expressed as

$$M_{g_d} = \begin{bmatrix} M_{11} & 4(-L_{p_1} + \ell)(L_{p_1} + \ell) & -4(L_{p_1} + \ell)(L_{p_2} + \ell) & 4(L_{p_1} + \ell)(-L_{p_2} + \ell) \\ -4(-L_{p_1} + \ell)(L_{p_1} + \ell) & M_{22} & -4(-L_{p_1} + \ell)(L_{p_2} + \ell) & 4(-L_{p_1} + \ell)(-L_{p_2} + \ell) \\ -4(L_{p_2} + \ell)(L_{p_1} + \ell) & 4(L_{p_2} + \ell)(-L_{p_1} + \ell) & M_{33} & 4(-L_{p_2} + \ell)(L_{p_2} + \ell) \\ -4(-L_{p_2} + \ell)(L_{p_1} + \ell) & 4(-L_{p_2} + \ell)(-L_{p_1} + \ell) & -4(-L_{p_2} + \ell)(L_{p_2} + \ell) & M_{44} \end{bmatrix} /(2i)^2. \quad (11)$$

$$\begin{aligned} M_{11} &= -4(2L_{p_1}^2 + L_{p_1}L_{p_2} + 4L_{p_1}\ell + 2L_{p_2}\ell + \ell^2) \\ M_{22} &= 4(2L_{p_1}^2 + L_{p_1}L_{p_2} - 4L_{p_1}\ell - 2L_{p_2}\ell + \ell^2) \\ M_{33} &= -4(L_{p_1}L_{p_2} + 2L_{p_1}\ell + 2L_{p_2}^2 + 4L_{p_2}\ell + \ell^2) \\ M_{44} &= 4(L_{p_1}L_{p_2} - 2L_{p_1}\ell + 2L_{p_2}^2 - 4L_{p_2}\ell + \ell^2) \end{aligned} \quad (12)$$

Here $g'_d = \hbar\omega_\beta \mathcal{G}'_d$ and $\mathcal{G}'_d = 2g_{d_{\sigma\chi}}(N/2\pi)$ is the drag factor. The eigenvalues of the above matrix $M$ [Eq. (7)] represent the sidemode frequencies of the BEC matter waves. Although in principle they can be obtained analytically, the expressions are cumbersome, and in practice we have calculated them numerically. The corresponding sidemode frequencies have been represented by dashed lines in the plots showing the power spectrum of the phase quadrature of the cavity output field in the main article.

## III. COUNTERSUPERFLOW INSTABILITY

Countersuperflow instability occurs when the relative velocity between two superfluids is large enough to lead to the transfer of momentum aimed at reducing their relative momentum while conserving the total momentum in the system [5]. The system stability phase diagram for various values of inter-species atomic interactions is shown in Fig. 2, by considering the eigenfrequencies of matrix $M$ [Eq. (7)] as functions of the parameters $L_{p_1}$ and $\ell$. In the

absence of interactions ($\mathcal{G}'_d = 0$) it can be seen from Fig. 2 (a) that the system is dynamically stable for all values of $L_{p_1}, \ell$ within the ranges considered. In the presence of interactions ($\mathcal{G}'_d \neq 0$), the frequencies are complex inside the red region of Fig. 2(b), which indicates that the condensate is dynamically unstable for those values of $L_{p_1}$ and $\ell$. In this case, for a given $\ell$, the condensates become dynamically unstable above a critical value of $L_{p_1}$. Rotation of the optical lattice results in the appearance of one more unstable branch, as shown in Fig. 2 (c). For higher values of inter-species interactions, the critical value of $L_{p_1}$ decreases, and the width of the unstable region increases. These findings are similar to the earlier analysis for only density-dependent atomic interactions [6].

## IV. DYNAMIC BRAGG SPECTROSCOPY: LATTICE ROTATION

In this section we demonstrate the dynamic Bragg spectroscopy method used by us to amplify the peak splitting in the cavity spectrum. Essentially, by rotating the weak optical lattice at angular frequency $\Omega'$ the separation between the first order Bragg diffraction peaks can be varied [7].

This effect can be seen in Fig.3, which shows cavity spectra for various values of $\Omega'$. There is no rotation ($\Omega' = 0$) in Fig.3(a), where the peak splitting $\omega_{2+} - \omega_-$ is not resolvable for our parameters. In contrast, Figs.3(b)-(e) show how the peak splitting becomes clearly and increasingly resolvable for increasing values of $\Omega'$.

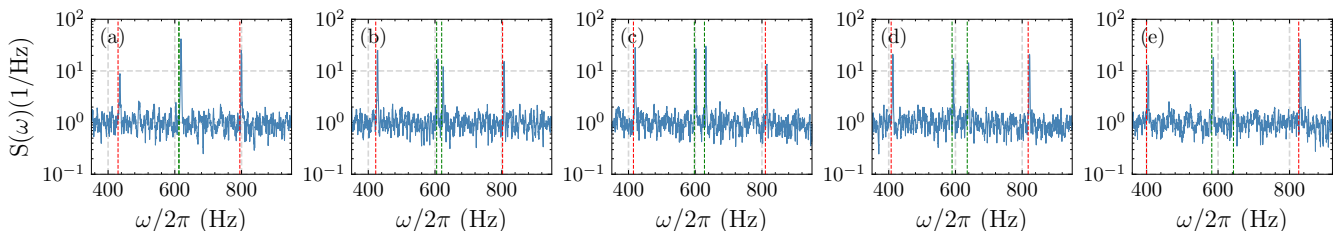

FIG. 3. Power spectra of the output phase quadrature of the cavity field for (a) $\Omega' = 0$, (b) $\Omega' = -0.25$, (c) $\Omega' = -0.5$, (d) $\Omega' = -0.75$, and (e) $\Omega' = -1$. the other parameters used are $L_{\psi_2}$ for $L_{p_1} = 3, L_{p_2} = 0$, $N = 1500$, $\mathcal{G}' = 0.8$, $\mathcal{G}'_d = 0.095$, $P_{\text{in}} = 0.4\,\text{pW}$. Here The red (green) dashed lines denote the analytical predictions for the frequencies corresponding to $L_{p_1} \pm 2\ell - \Omega'/2$ ($L_{p_2} \pm 2\ell - \Omega'/2$) sidemodes, obtained from the BdG analysis.

## V. CALCULATION OF THE DRAG FACTOR

In a two-component BEC having equal component densities $\rho_1 = \rho_2$ and the drag density $\rho_d$, the drag factor is related to the intra-species and inter-species atomic interactions $g_{\sigma\sigma}$ and $g_{\sigma\sigma'}$, respectively, and is expressed analytically using Bogoliubov theory as [8]

$$\mathcal{G}'_d = 2\frac{\rho_d}{\rho_1} = \frac{16\eta^2}{3\pi} \frac{1}{\left(\sqrt{2(\gamma+\eta)} + \sqrt{2(\gamma-\eta)}\right)^3}, \qquad (13)$$

where $\rho_d/\rho_1 = g_{d_{\sigma_\chi}}\rho_1$. Here $\gamma = 2g_{\sigma\sigma}m/\hbar^2 n$ and $\eta = 2g_{\sigma\sigma'}m/\hbar^2 n$ are the dimensionless atomic interaction strengths, $n$ is the total BEC density defined as $n = (2N)/(2\pi R)$, $g_{\sigma\chi} = 2\hbar\omega_\rho a_{s_{\sigma\chi}}/R$ denotes the reduced atomic interactions. After substituting the values provided in the main article, we get $g_{\sigma\sigma} = 1.16 \times 10^{-35}$. Considering $g_{\sigma\sigma'} = \mathcal{G}' g_{\sigma\sigma}$, for $\mathcal{G}' = \{0.1, 0.2, 0.3, 0.4, 0.5, 0.6, 0.7, 0.8, 0.9\}$, we obtain $\mathcal{G}'_d = \{0.001, 0.004, 0.01, 0.018, 0.029, 0.045, 0.066, 0.095, 0.141\}$.

## VI. POWER SPECTRUM OF ANGULAR MOMENTUM OF THE TWO COMPONENTS

In this section, we have calculated the Fourier transform of the time series of the angular momentum per atom of both superfluid components, which yields the characteristic frequencies of the angular momentum variation, and compared it with the cavity output spectrum. The power spectrum of the phase quadrature of the cavity output field for $L_{p_1} = 3$, $L_{p_2} = 0$, and $\mathcal{G}'_d = 0.095\,(\mathcal{G}' = 0.8)$ is shown in Fig. 4 (a). As explained in the main article, we see four peaks, corresponding to the $L_{p_1} \pm 2\ell - \Omega'/2$ and $L_{p_2} \pm 2\ell - \Omega'/2$ sidemodes at the locations predicted by the Bogoliubov analysis.

The Fourier transformations of the angular momenta per atom $L_{\psi_1}$ are shown in Fig. 4 (b). In this plot, the dominant peaks correspond to $L_{p1} \pm 2\ell - \Omega'/2$ due to the initial motion of the $\psi_1$ component. In addition, peaks

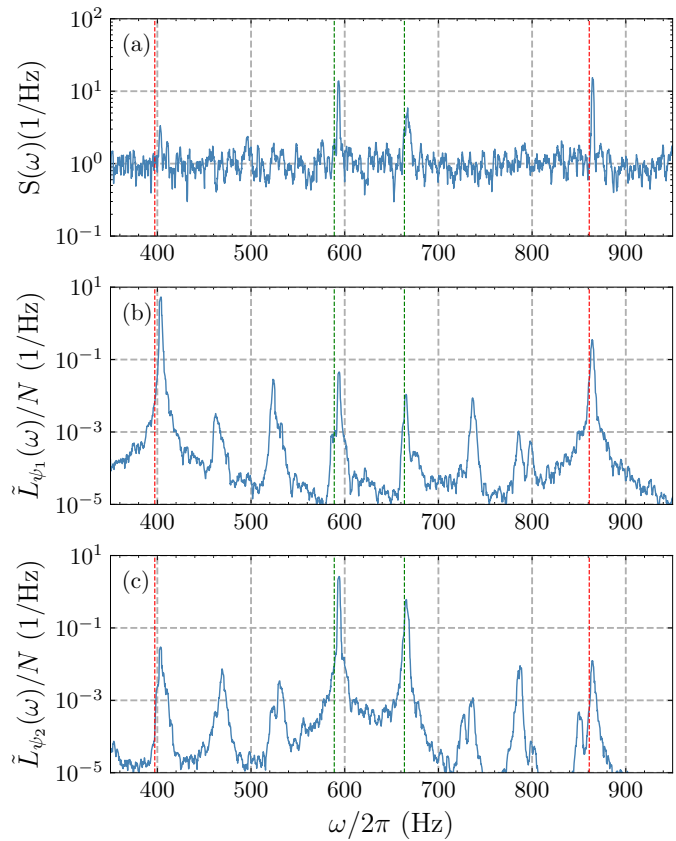

FIG. 4. (a) The power spectrum of the phase quadrature of the cavity output field, (b) the power spectrum of $L_{\psi_1}$, and (c) the power spectrum of $L_{\psi_2}$ for $L_{p_1} = 3, L_{p_2} = 0, \Omega' = -1, N = 1500, \mathcal{G}' = 0.8, \mathcal{G}'_d = 0.095, P_{\text{in}} = 0.4\,\text{pW}$. The red (green) dashed lines denote the analytical predictions for the frequencies corresponding to $L_{p_1} \pm 2\ell - \Omega'/2$ ($L_{p_2} \pm 2\ell - \Omega'/2$) side modes, and were obtained from the Bogoliubov analysis.

corresponding to the winding number of the other component, namely $\psi_2$, also appear, signifying the presence of angular momentum exchange. The contribution of the current-interaction-driven AB effect is larger than that of the countersuperflow instability, making the frequency splitting resolvable. The other (unlabeled) peaks in the spectrum appear due to the involvement of the neighboring modes of $L_{p_1} \pm 2\ell - \Omega'/2$ and $L_{p_2} \pm 2\ell - \Omega'/2$ in the dynamics of the transition from the uniform density state to the density modulated state. The Fourier transformation of the angular momentum per atom $L_{\psi_2}$ is likewise shown in Fig. 4 (c) and has a similar explanation.

## VII. EFFECT OF INPUT OPTICAL POWER ON THE CAVITY OUTPUT SPECTRA

The power spectrum of the phase quadrature of the cavity output field normalized to the steady state photon number for various input optical power values is shown in Fig. 5.

Here we have focused on the $\omega_{2+}$ peak corresponding to the initially non-rotating condensate ($L_{p_2} \pm 2\ell - \Omega'/2$), which is the object being measured in our protocol. Increasing the input power initially improves signal-to-noise, decreasing the optical shot noise. After a certain value of input power ($P_{\text{in}} = 0.075$ pW), the noise level begins to increase due to back action noise [2]. This figure indicates the existence of a standard quantum limit, and hence the optimum measurement power value of $P_{\text{in}} = 0.075$ pW, where the optical shot noise and back action noise due to the radiation pressure force are balanced [2, 9].

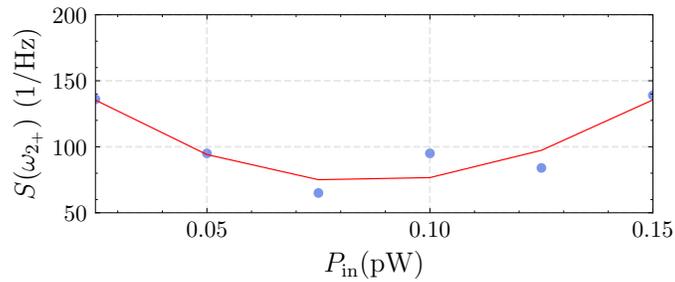

FIG. 5. Variation of the peak height at $\omega_{2+}$ with input optical power. The other parameters used are the same as in Fig. 4. The red line denotes a third-order polynomial curve that fits the data points.

## VIII. AB DRAG EFFECT IN A POPULATION IMBALANCED BEC

In this section, we investigate a spinor BEC with unequal particle numbers ($N_1 \neq N_2$), a case that is more experimentally accessible than the balanced spinor. We show the condensate density, phase profiles, the corresponding matter wave OAM distributions, the angular momentum dynamics, and the power spectra of the output phase quadrature of the cavity field in Fig. 6. Remarkably, regardless of the unequal particle numbers, our proposed method is able to determine the drag factor from the cavity output spectra, providing a robust approach to detect the AB drag effect in an experimentally realistic setup.
---


\fi

\end{document}